\documentstyle [aps,preprint,epic,eepic]{revtex}
\tightenlines
\begin{document}
\def\fnote#1#2{
\begingroup\def\thefootnote{#1}\footnote{#2}\addtocounter{footnote}{-1}
\endgroup}
\def\dslash{\not{\hbox{\kern-2pt $\partial$}}}
\def\eslash{\not{\hbox{\kern-2pt $\epsilon$}}}
\def\Dslash{\not{\hbox{\kern-4pt $D$}}}
\def\Aslash{\not{\hbox{\kern-4pt $A$}}}
\def\Qslash{\not{\hbox{\kern-4pt $Q$}}}
\def\Wslash{\not{\hbox{\kern-4pt $W$}}}
\def\pslash{\not{\hbox{\kern-2.3pt $p$}}}
\def\kslash{\not{\hbox{\kern-2.3pt $k$}}}
\def\qslash{\not{\hbox{\kern-2.3pt $q$}}}
\def\np#1{{\rm Nucl.~Phys.~\bf B#1}}
\def\pl#1{{\rm Phys.~Lett.~\bf B#1}}
\def\mpl#1{{\rm Mod.~Phys.~Lett.~\bf A#1}}
\def\pr#1{{\rm Phys.~Rev.~\bf D#1}}
\def\prl#1{{\rm Phys.~Rev.~Lett.~\bf #1}}
\def\cpc#1{{\rm Comp.~Phys.~Comm.~\bf #1}}
\def\cmp#1{{\rm Commun.~Math.~Phys.~\bf #1}}
\def\anp#1{{\rm Ann.~Phys.~(NY) \bf #1}}
\def\etal{{\em et al.}}
\def\half{{\textstyle{1\over2}}}
\def\be{\begin{equation}}
\def\ee{\end{equation}}
\def\ba{\begin{array}}
\def\ea{\end{array}}
\def\tr{{\rm tr}}
\def\Tr{{\rm Tr}}
\title{Temperature of D3-branes off extremality
\thanks{Research
supported by the DoE under grant DE--FG05--91ER40627.}}
\author{
Suphot Musiri\fnote{\dagger}{E-mail: \tt smusiri@utk.edu} and George Siopsis
\fnote{\ddagger}{E-mail: \tt gsiopsis@utk.edu}}
\address{Department of Physics and Astronomy, \\
The University of Tennessee, Knoxville, TN 37996--1200.\\
}
\date{March 2000}
\preprint{UTHET--00--0301}
\maketitle
\begin{abstract}
We discuss non-extremal rotating D3-branes. We solve the wave equation
for scalars in the supergravity background of certain distributions of
branes and compute the absorption coefficients. The form of these coefficients
is similar to the gray-body factors associated with black-hole scattering.
They are given in terms of two different temperature parameters, indicating
that fields (open string modes) do not remain in thermal
equilibrium as we move off extremality. This should shed some light on the origin of the
disagreement between the supergravity and conformal field theory results on
the free energy of a system of non-coincident D-branes. \end{abstract}
\renewcommand\thepage{}\newpage\pagenumbering{arabic} 
A useful tool in the study of the AdS/CFT correspondence~\cite{bib1,bib2,bib3}
is the investigation of interactions of branes
with external probes~\cite{bib8a,bib8b,bib9,bib9a,bibgs}. In particular, the 
absorption cross-section for a scalar
in an AdS background has been shown to agree with the one obtained from
superconformal field theory. This agreement has been shown to be exact in the
low frequency limit and for all partial waves of the scalar
field~\cite{bib8a}. Extensions to higher frequencies have also been 
considered~\cite{bib9}. In the more
general case of non-coincident D-branes, such calculations are considerably
more involved, because the wave equation becomes non-separable~\cite{bibgs}.

In ref.~\cite{bibgs}, the absorption of scalars by a continuous
distribution of
D3-branes was studied.
The wave equation was solved for arbitrary
partial waves and a large set of supergravity backgrounds
corresponding to various D-brane
distributions in the extremal limit. In general, the waves become singular
at the positions of the D-branes.
It was found that the absorption coefficients exhibit a universal
behavior similar to the form of the gray-body factors in the case of
black-hole scattering~\cite{bib9b}. The study of a general class of
supergravity backgrounds is necessary if one's goal is to understand
quantum gravity. Such studies have already appeared in the literature, starting 
with linearized
perturbations around the special solution~\cite{bib7ab}
and including solutions of the full non-linear
field equations~\cite{bib7a,bib7aa}. Of particular importance are solutions that
represent
a collection of branes (multi-center). They correspond to a broken phase of the
super Yang-Mills theory where certain operators develop non-vanishing
vacuum expectation values. This is the Coulomb branch of the gauge theory,
because the remaining massless bosons mediate long-range Coulomb
interactions. Superconformal symmetry is broken. The space of these solutions
is the moduli space of the Yang-Mills theory. One recovers the AdS limit
by approaching the origin of moduli space.

To understand the thermodynamic properties of gauge theories, such as phase 
transitions~\cite{Gubser,bib8,bibcai}, one needs to consider non-extremal solutions of the
supergravity field equations which have finite temperature. As a step in this
direction, we study the scattering of scalars by brane configurations off
extremality. We calculate the absorption coefficients and show that, as we
move away from extremality, the degrees of freedom do not seem to remain in
equilibrium. Instead, they
evolve into separate systems at two distinct temperatures. This
should shed some light on the disagreement between the supergravity and
super-Yang-Mills calculations of the free-energy
density~\cite{bibgkp,Klebanov,Fotopoulos,Mozo,Rey,biboh}. The two resultant
expressions disagree by a factor of 3/4, indicating that the discrepancy is
related to an incorrect account of the number of degrees of freedom. 


The metric for a general distribution of spinning D3-branes in ten dimensions
can be written in terms of four parameters~\cite{bib7aa,bib7b,bib7c}, the moduli 
$\ell_i$
($i=1,2,3$), which are the angular momentum quantum numbers representing
rotation around
axes in three distinct planes, respectively, in the six-dimensional transverse 
space, and $r_0$, which is related to the position of the horizon. The limit
$r_0\to 0$ is the extremal limit. When the moduli $\ell_i$ ($i=1,2,3$) vanish,
$r_0$ is the position of the horizon.

To simplify the calculations, we will restrict attention to the special case
\be
\ell_2 = \ell_3 = 0
\ee
The more general case is expected to lead to qualitatively similar results.
The metric in this special case is given by
$$ds^2 = {1\over\sqrt H} \left( - (1- r_0^4/\zeta r^4) dt^2 + 
dx_1^2+dx_2^2+dx_3^2 \right)
+ \sqrt H \zeta {dr^2\over \lambda - r_0^4/r^4}$$
$$- {2r_0^4\cosh\gamma\over \zeta r^4\sqrt H} \; \ell_1\sin^2\theta d\phi_1\; dt
+ {r_0^4\over \zeta r^4 \sqrt H}\; \ell_1^2\sin^4\theta d\phi_1^2$$
\be\label{miden1}
+ \sqrt H r^2 \left[ \zeta d\theta^2 + \lambda \sin^2\theta d\phi_1^2
+\cos^2\theta (d\psi^2
+ \sin^2\psi d\phi_2^2
+ \cos^2\psi d\phi_3^2) \right]
\ee
where
\be\label{dekatria}
H = 1+\; {r_0^4\sinh^2\gamma\over \zeta r^4} \quad\quad 
\lambda = 1+{\ell_1^2\over r^2}
\quad\quad \zeta = 1+ {\ell_1^2\cos^2\theta \over r^2}
\ee
Notice that the various functions, $H, \lambda, \zeta$, comprising the metric 
tensor, are functions of $r, \theta$
only. 
The horizon is one of the roots of $\lambda - r_0^4/r^4 =0$,
\be\label{horr}
r_H^2 = {1\over 2} \; \left( \sqrt{\ell_1^4 + 4r_0^4} - \ell_1^2\right)
\ee
The other solution of $\lambda - r_0^4/r^4 =0$ is $-r_+^2$, where
\be
r_+^2 = {1\over 2} \; \left( \sqrt{\ell_1^4 + 4r_0^4} + \ell_1^2\right)
\ee
It is convenient to introduce the parameter $\Delta$:
\be\label{paradelta}
\Delta = {r_H^2\over r_+^2}
\ee
The limit $\Delta\to 0$ corresponds to $r_0 \to 0$ (extremal limit). The
limit $\Delta\to 1$ corresponds to $\ell_1\to 0$. We will study
both these limits. The other scale in the metric is the
D-brane charge
\be\label{dbch}
R^4 = {1\over 2} \; r_0^4\; \sinh (2\gamma)
\ee
and will be assumed to be much larger than either
$\ell_1$ or $r_0$:
\be
R \gg \ell_1 \;,\; r_0 \;,\; r_H \;,\; r_+
\ee
The energy, angular momentum, and entropy densities, Hawking temperature
and angular velocity are, respectively,
$$\epsilon = {1\over 4G}\; r_0^4 (4\cosh^2\gamma - 4\cosh\gamma\sinh\gamma +1)
\quad\quad j = {1\over 2G} r_0^4 \ell_1 \cosh\gamma$$
\be\label{miscv}
s = {\pi\over G} \; r_0^4 r_H \cosh\gamma \quad T_H = {r_H\over 2\pi r_0^4
\cosh\gamma} \; \sqrt{\ell_1^4 + 4r_0^4}\quad\quad
\Omega_H = {\ell_1 r_H^2
\over r_0^4 \cosh\gamma}
\ee
where $G$ is a constant related to Newston's constant. These quantities
obey the thermodynamic relation
\be
T_H ds = d\epsilon - \Omega_H dj
\ee
In the extremal limit, the horizon shrinks to zero ($r_0\to 0$) and also
$\gamma \to\infty$, so that the charge $R^4$ remains finite. The angular 
momentum
also vanishes and we obtain a static configuration. Nevertheless,
these configurations are
still described by the angular momentum quantum number. Notice that the
temperature parameter remains finite, even though the entropy and energy
densities vanish. It should be pointed out that the physical meaning of this
temperature parameter is not obvious, because the curvature has a signularity
in this limit.
The metric in the extremal limit becomes
$$ds^2 = {1\over\sqrt H}
\left( - dt^2 + dx_1^2+dx_2^2+dx_3^2 \right) + \sqrt H \zeta {dr^2\over
\lambda}$$
\be
+ \sqrt H r^2 \left[ \zeta d\theta^2 + \lambda \sin^2\theta
d\phi_1^2 + \cos^2\theta \Big(d\psi^2 + \sin^2\psi d\phi_2^2 + \cos^2\psi
d\phi_3^2\Big) \right]
\ee 
where
\be
H = 1+ {R^4\over \zeta r^4}
\ee
and the other functions, $\lambda, \zeta$ are still given by 
Eq.~(\ref{dekatria}). It can be shown that this metric is equivalent to the
multi-center form
\be\label{deka}
ds^2 = {1\over\sqrt H} \; \left( -dt^2+dx_1^2+dx_2^2+dx_3^2 \right)
+ \sqrt H\; \left( dy_1^2+\dots + dy_6^2  \right)
\ee
with the harmonic function $H$ given by
\be\label{dodeka}
H = 1 + R^4\; \int d^6 y' {\sigma ({\bf y'})\over |{\bf y} - {\bf y'}|^4}
\;, \quad\quad \int d^6 y' \sigma ({\bf y'}) = 1
\ee
where $\sigma ({\bf y})$ describes the distribution of branes,
through the
following transformation~\cite{bib7aa,bib11}
$$y_1 = \sqrt{r^2 + \ell_1^2} \sin\theta \cos\phi_1$$
$$y_2 = \sqrt{r^2 + \ell_1^2} \sin\theta \sin\phi_1$$
$$y_3 = \cos\theta \sin\psi \cos\phi_2$$
$$y_4 = \cos\theta \sin\psi \sin\phi_2$$
$$y_5 = \cos\theta \cos\psi \cos\phi_3$$
\be\label{cootr}
y_6 = \cos\theta \cos\psi \sin\phi_3
\ee
To see this, note that
the harmonic function $H$ (Eq.~(\ref{dodeka})) can be written as
\be
H = 1+ {R^4\over \zeta r^4} = 1+ {R^4\over (r^2+\ell_1^2\cos^2\theta)r^2}
\ee
The D-branes are in the region bounded by the $r=0$ surface, which is a disk of
radius $\ell_1$ in the plane $y_3=\dots = y_6=0$ (because of Eq.~(\ref{cootr})).
Define
\be
y_{||}^2 = y_1^2+y_2^2 = (r^2+\ell_1^2) \sin^2\theta \;, \quad\quad
y_\perp^2 = y_3^2+\dots + y_6^2 = r^2\cos^2\theta
\ee
in terms of
which $H$ becomes
\be
H \approx 1+ {R^4\over \ell_1^2 y_\perp^2}
\ee
as $r\to 0$. The density of D-branes is therefore independent of $y_{||}$
and $\sigma = {1\over \pi\ell_1^2}$, {\em i.e.,} the D-branes are uniformly 
distributed on a disk of radius $\ell_1$ in the $y_{||}=0$ plane.

The ten-dimensional wave equation for a scalar field,
\be
\partial_A \sqrt{-g} g^{AB} \partial_B \Phi = 0
\ee
becomes separable for fields that are independent of the angular variables 
$\psi$, $\phi_i$
($i=1,2,3$).
For simplicity, we will further assume that there is no three-dimensional 
spatial dependence.
For a field of momentum $k_\mu = (\omega \;,\; \vec 0)$,
\be
\Phi (x^\mu\, ;\, r\,,\, \theta) = e^{i\omega t}\; \Psi (r\,,\, \theta)
\ee
after some algebra, we obtain
\be
{1\over r^3} \partial_r \left((\lambda -r_0^4/r^4)
r^5 \partial_r \Psi\right)
+ \omega^2 r^2\Psi + {\omega^2\lambda r_0^4\cosh^2\gamma\over r^2
(\lambda - r_0^4/ r^4
)} \Psi - (\hat L^2 -\omega^2\ell_1^2\cos^2\theta) \Psi =0
\ee
We will solve this equation in the limit where the mass is small
compared with the AdS curvature, and the angular momentum is also small,
\be
R\;,\; \ell_1 \ll 1/\omega
\ee
In this limit, the term proportional to the angular momentum can
be dropped. Indeed, its contribution is
$\omega^2\ell_1^2  \ll 1$.
Therefore, it is small compared to the angular momentum ($\hat L^2$ term)
contribution.
The wave equation becomes
\be
{1\over r^3} \partial_r \left((\lambda -r_0^4/r^4)
r^5 \partial_r \Psi\right)
+ \omega^2 r^2\Psi + {\omega^2\lambda r_0^4\cosh^2\gamma\over r^2
(\lambda - r_0^4/ r^4
)} \Psi - \hat L^2 \Psi =0
\ee
The eigenvalues of $\hat L^2$ are $j(j+4)$. Therefore, the radial part
of the wave equation is
\be
{1\over r^3} \partial_r \left((\lambda -r_0^4/r^4)
r^5 \partial_r \Psi\right)
+ \omega^2 r^2\Psi + {\omega^2\lambda r_0^4\cosh^2\gamma\over r^2
(\lambda - r_0^4/ r^4
)} \Psi - j(j+4) \Psi =0
\ee
We will solve this equation in two regimes, $r \gg \omega
R^2$ and $r\ll 1/\omega$,
and then match the respective expressions asymptotically.

For $r \gg \omega R^2$, we obtain
\be
{1\over r^3} \left( r^5 \Psi'\right)'
+ \omega^2r^2\Psi -  j(j+4)\Psi =0
\ee
whose solution is
\be
\Psi = {1\over r^2} \; J_{j+2} (\omega r)
\ee
where we dropped the solution which is not regular at small $r$. The 
normalization is arbitrary, since we only care about ratios of fluxes.
At small $r$, the solution
behaves as
\be\label{tria}
\Psi \sim {\omega^2\over 4(j+2)!} \; \left( {\omega r\over 2} \right)^j
\ee
In the regime of small $r$ ($\omega r \ll 1$), the wave equation becomes
\be\label{duo1}
{1\over r^3} \partial_r \left((\lambda -r_0^4/r^4)
r^5 \partial_r \Psi\right)
 + {\omega^2\lambda r_0^4\cosh^2\gamma\over r^2
(\lambda - r_0^4/ r^4
)} \Psi - j(j+4) \Psi =0
\ee
We will solve this equation at extremality as
well as away from it. We will obtain expressions for the absorption
coefficients indicating that two distinct temperatures enter the
problem and flow to different values as one varies the parameter $\Delta$
(Eq.~(\ref{paradelta})). The two temperatures coincide at the extremal
limit $\Delta =0$.

\noindent\underline{\sl I. The end-point $\Delta = 0$}

We start with the extremal limit where $r_0\to 0$ and the horizon shrinks
to zero. We will follow the discussion of ref.~\cite{bibgs}.
As explained above, the extremal limit represents branes
which are uniformly distributed on
a two-dimensional disk of radius $\ell_1$ in the transverse 
space~\cite{bib7a,bib7aa}.
This is a Bose-Einstein condensate. It has zero entropy because it has
settled into its ground state. However, the temperature remains finite
implying that the temperature parameter lacks a standard physical
interpretation. Nevertheless, we shall continue to refer to it as the
temperature of the system, because it possesses the usual thermodynamic
properties.
      
The wave equation~(\ref{duo1}) becomes
\be\label{duo2}
{1\over r^3} \left(\lambda r^5 \Psi'\right)'
 + {R^4\omega^2\over r^2} \Psi -  j(j+4)\Psi =0
\ee
To solve this equation, change variables to $u = 1/\lambda$. Then
\be
(1-u) u^2 {d^2\Psi\over du^2} + u(2-u) {d\Psi\over du} + {\omega^2R^4\over 
4\ell_1^2} \Psi - {j(j+4)u\over
4 (1-u)} \Psi =0
\ee
Next, we need to control the behavior at the singular points $u=0,1$.
As $u\to 0$, we obtain
\be\label{tria1}
u^2 {d^2\Psi\over du^2} + 2u {d\Psi\over du} +
{\omega^2R^4\over 4\ell_1^2} \Psi - {j(j+4)\over 4} u \Psi = 0
\ee
Assuming $\Psi \sim u^a$, we obtain
\be\label{ena}
a = - {1\over 2} + {1\over 2} \sqrt{1-{\omega^2R^4\over \ell_1^2}}
= - {1\over 2} + i\kappa\;, \quad\quad
\kappa = {1\over 2} \sqrt{{\omega^2R^4\over \ell_1^2} -1} \approx
{\omega R^2\over 2\ell_1} = {\omega\over 4\pi T_H}
\ee
where $T_H = {\ell_1\over 2\pi R^2}$ is the Hawking temperature.
 Let us point
out again that, this being an extremal configuration,  one would expect the
temperature to vanish (just as the entropy and energy do). This does not
happen owing to the curvature signularity at the horizon. Therefore, this
temperature parameter lacks a standard physical interpretation. Nevertheless,
it possesses standard thermodynamic properties.

As $u\to 1$, we
obtain 
\be (1-u)^2 {d^2\Psi\over du^2} + (1-u) {d\Psi\over du} +
{\omega^2R^4\over 4\ell_1^2} \; (1-u)\Psi - {j(j+4)\over 4} \Psi = 0
\ee
Assuming $\Psi \sim (1-u)^b$, we obtain
$b = {j+4\over 2}$.
Now set
\be\label{tes1}
\Psi = A u^{-1/2 + i\kappa} \; (1-u)^{j/2+2} f(u)
\ee
Eq.~(\ref{tria1}) becomes
\be
(1-u)u {d^2f\over du^2} + [1+2i\kappa-(j+4+2i\kappa)u] {df\over du} -
{(j+3+2i\kappa)^2\over 4} \; f = 0
\ee
whose solution is the hypergeometric function
\be\label{pente1}
f(u) = F\left( (j+3)/2+i\kappa\; ,\; (j+3)/2+i\kappa\; ;\; 1+2i\kappa\; ;\; u
\right)
\ee
To obtain the behavior of $\Psi$ for large $r$, note that
\be\label{exi1}
F\left( (j+3)/2+i\kappa\; ,\; (j+3)/2+i\kappa\; ;\; 1+2i\kappa\; ;\; u
\right) = {\Gamma (1+2i\kappa)
\Gamma(j+2)\over (\Gamma((j+3)/2+i\kappa))^2}
 {1\over (1-u)^{j+2}} +\dots
\ee
where the dots represent terms that are regular in $1-u$.
Therefore, using Eqs.~(\ref{tes1}), (\ref{pente1}) and (\ref{exi1}), we arrive 
at
\be
\Psi \approx A \; {\Gamma (1+2i\kappa)
\Gamma(j+2)\over (\Gamma((j+3)/2+i\kappa))^2} \; \left( {r\over \ell_1}\right)^j
\ee 
Comparing with the asymptotic form~(\ref{tria}), we obtain
\be\label{exi}
A = {(\Gamma((j+3)/2+i\kappa))^2\over \Gamma (1+2i\kappa)}\;
{\omega^{j+2}\ell_1^j\over 2^{j+2}(j+1)!(j+2)!}
\ee
In the small $r$ limit, we have $u\approx r^2/\ell_1^2$ and Eq.~(\ref{tes1}) 
reads
\be
\Psi \approx A \left( {r\over \ell_1} \right)^{-1+2i\kappa}
\ee
The absorption coefficient, which is the ratio of the incoming flux
at $r\to 0$ to the incoming flux at $r\to \infty$, is
\be\label{epta1}
{\cal P} = {\Im (\lambda r^5 \Psi^\ast \Psi')|_{r\to 0}\over \Im (r^5 \Psi^\ast 
\Psi')|_{r\to\infty}}
= 4\pi \kappa \ell_1^4 |A|^2
= 4\pi \kappa \; {|\Gamma((j+3)/2+i\kappa)|^4\over |\Gamma (1+2i\kappa)|^2}\;
{\omega^{2j+4}\ell_1^{2j+4}\over 4^{j+2}((j+1)!(j+2)!)^2}
\ee
This is of the same form as the grey-body factors obtained in black-hole 
scattering~\cite{bib9b} for large $j$.
Indeed, comparing
\be\label{absco}
{\cal P} \sim |\Gamma((j+3)/2+i\kappa)|^4 = \left|\Gamma\left(
{j+3\over 2}+i\; {\omega\over 4\pi T_H}\right)\right|^4
\ee
with the general form of a grey-body factor~\cite{bib9b},
\be
{\cal P}_{b.h.} \sim \left|\Gamma\left(
{j+2\over 2}+i\; {\omega\over 4\pi T_+}\right)\right|^2\left|\Gamma\left(
{j+2\over 2}+i\; {\omega\over 4\pi T_-}\right)\right|^2
\ee
we see that we get contributions from all fields at
the same temperature,
\be T_+=T_-=T_H\ee
For completeness, we mention that
Eq.~(\ref{epta1}) also reproduces results derived earlier in the AdS (zero
temperature) limit. Indeed,
in the small temperature limit, we have $\kappa \to \infty$ and
\be
A \approx 
\sqrt\pi\; {i^{j+2} R^{2j+4}\omega^{2j+4}
\over 4^{j+2+i\kappa/2}\ell_1^2\; (j+1)!(j+2)!} \; {\Gamma (1/2+i\kappa)\over
\Gamma (1+i\kappa)}
\ee
for even $j$, where we used the Gamma function identities
\be\label{okto}
\Gamma (2x) = {1\over \sqrt{2\pi}} \; 2^{2x-1/2} \Gamma(x)\Gamma(x+1/2)
\quad\quad
\Gamma(x+1) = x \Gamma(x)
\ee
Since also $|\Gamma ({1\over 2} + i\kappa)|^2 = \pi / \cosh (\pi\kappa)$ and
$|\Gamma (1 + i\kappa)|^2 = \pi\kappa / \sinh (\pi\kappa)$, we obtain
\be
|A|^2 \approx { \pi R^{4j+8}\omega^{4j+8}
\over 4^{2j+4}\kappa \ell_1^4\; ((j+1)!(j+2)!)^2}
\ee
and so the absorption coefficient~(\ref{epta1}) becomes
\be\label{okto1}
{\cal P} = 4\pi \kappa \ell_1^4 |A|^2
\approx { \pi^2 R^{4j+8}\omega^{4j+8} \over 4^{2j+3} \; ((j+1)!(j+2)!)^2}
\ee
in agreement with earlier results~\cite{bib8a}.

Thus, we have established that the absorption coefficients (Eq.~(\ref{absco}))
are given in terms of a single temperature parameter (the Hawking temperature)
in the extremal limit. Of course, the system has settled into its ground state
and this result does not have an obvious physical interpretation in terms of
thermal equilibrium. Next, we wish to move away from extremality, where the
horizon as well the physical temperature become finite. It is hard to find an analytic form of the solution to the wave equation~(\ref{duo1})
for a general finite parameter $\Delta$. We have managed to derive the solution
in closed form in the limit of small $\Delta$ and
at $\Delta =1$.

\noindent\underline{\sl II. Approaching the limit $\Delta\to 0$}

Away from extremality, the horizon is finite and we first need to isolate the
behavior of the wavefunction near the horizon.
Let the asymptotic form of the wavefunction be
\be
\Psi \sim \left( 1 - {r_H^2\over r^2} \right)^b
\ee
where $r_H$ is the radius of the horizon (Eq.~(\ref{horr})).
Substituting into the wave equation~(\ref{duo1}), we obtain
\be
b = {i \omega \cosh\gamma r_0^4\over 2r_H\sqrt{\ell_1^4+4r_0^4}}
= {i\omega\over 4\pi T_H}
\ee
where $T_H$ is the Hawking temperature (Eq.~(\ref{miscv})).
The wavefunction for $r \ge r_H$ can then be written as
\be
\Psi = \left( 1 - {r_H^2\over r^2} \right)^b \; f(r)
\ee
where the function $f(r)$ is regular at the horizon ($r=r_H$).
After some algebra, we obtain
$$
r^2(\lambda - r_0^4/r^4) f'' + \left[ 2 (2b+1) {r_H^2\over r^2}
\left( 1+{r_+^2\over r^2} \right) + \left( 5+3{r_+^2\over r^2} \right)
\left( 1-{r_H^2\over r^2} \right) \right] rf'$$
\be\label{dekatrianew}
+ \left[ 4{r_H^2\over r^2} - {4b^2\over r^2}
\left( {(r_+^2+r_H^2)r_H^2\over r_+^2(1+r_+^2/r^2)} +r_+^2 + 
r_H^2(1+r_+^2/r^2)\right) - j(j+4)\right] f =0
\ee
In the limit $\Delta \to 0$, we have $r_+ \gg r_H$ ({\em 
cf.}~Eq.~(\ref{paradelta})).
So we can
solve Eq.~(\ref{dekatrianew}) in the $r\gg r_H$ regime and then take
the limit $r\to r_H$. This is guaranteed to be finite, because the
singularity has already been removed. We obtain
\be\label{dekatria99}
r^2\left( 1+{r_+^2\over r^2}\right) f'' +
\left( 5+3{r_+^2\over r^2} \right)
 rf'+ \left[  - {4b^2r_+^2\over r^2}
 - j(j+4)\right] f   =0
\ee
which is of the same form as~(\ref{duo2}). Therefore, the solution is
\be
f(u) = A u^{-1/2+b} (1-u)^{j/2+2} F((j+3)/2+b\;,\;
(j+3)/2+b\;,\; 1+2b \;,\; u)
\ee
where
\be
u = \left( 1+ {r_+^2\over r^2} \right)^{-1}\quad\quad
A = {(\Gamma((j+3)/2+b))^2\over \Gamma (1+2b)}\;
{\omega^{j+2} r_+^j\over 2^{j+2}(j+1)!(j+2)!}
\ee
At the horizon, $r = r_H$, we have $u = u_H = \Delta / (1+\Delta)$ and
$$
f(u_H) = A
\; u_H^{-1/2+b} (1-u_H)^{j/2+2} F((j+3)/2+b\;,\;
(j+3)/2+b\;,\; 1+2b \;,\; u_H)
$$
\be\label{fuh}
= A \Delta^{-1/2+b} \; \left( 1 +
{((j+3)/2+b)((j+1)/2-b)\over 1+2b} \; \Delta + o(\Delta^2) \right)
\ee
For large $j$ and $|b| \gg j$, we can use the Gamma function identities
\be\label{gfi}
\Gamma (2x) = {1\over \sqrt{2\pi}} \; 2^{2x-1/2} \; \Gamma (x) \Gamma(x+1/2)
\quad,\quad \Gamma(x+1) = x\Gamma(x)
\ee
to show that~(\ref{fuh})
takes on the asymptotic form
\be
f(u_H) = {\omega^{j+2} r_+^j\Delta^{-1/2+b}
(1+\Delta)^{(j-b)/2}\over 2^{j+2}(j+1)!(j+2)!}\;
{(\Gamma (1+b))^2\Gamma((j+3)/2+b_+)\Gamma((j+3)/2+b_-)\over \Gamma(1+b_+)
\Gamma(1+b_-)\Gamma (1+2b)}
\ee
where
\be
b_\pm = b\left( 1 \pm \sqrt{\Delta\over 2} \right)
\ee
Define temperature parameters
\be\label{maineq}
T_\pm = {T_H\over 1\pm \sqrt{\Delta\over 2}}
\ee
They satisfy
\be
{1\over T_+} + {1\over T_-} = {2\over T_H}
\ee
and $T_\pm \to T_H$ as $\Delta \to 0$. Thus, in the extremal limit ($\Delta
=0$) we recover thermal equilibrium at temperature $T=T_H$.

The absorption coefficient can be written as
\be\label{absne}
{\cal P} \sim \left| \Gamma((j+3)/2+i\omega/(4\pi T_+))
\Gamma((j+3)/2+i\omega/(4\pi T_-))\right|^2
\ee
indicating that the fields are not in thermal equilibrium away from extremality.
As the size of the horizon, $r_H$, increases ({\em i.e.}~as $\Delta$
increases), the temperature of some of the fields decreases whereas the
rest of the fields become hotter. It should be noted that Eq.~(\ref{maineq})
is only valid for small $\Delta$, so both temperature parameters $T_\pm$
are positive. It would be interesting to generalize Eq.~(\ref{maineq}) to
finite values of the parameter $\Delta$, but we have been unable to obtain
an analytic expression for the temperatures. Further support to the existence
of two distinct temperatures is obtained by examining the other end of the
spectrum of $\Delta$ ($\Delta =1$) which corresponds to the maximum value
of the radius of the horizon $r_H = r_0$.

\noindent\underline{\sl III. The other end-point, $\Delta = 1$}

The other end of the range of the parameter $\Delta$ ($\Delta = 1$,
Eq.~(\ref{paradelta}))
corresponds to a maximum radius of the horizon ($r_H = r_0$) and the vanishing 
of
the angular momentum quantum number $\ell_1 = 0$ (the other two angular momentum
quantum numbers have been set to zero from the outset). The wave 
equation~(\ref{duo1}) becomes
\be\label{eq99}
{1\over r^3} \partial_r \left((1 -r_0^4/r^4)
r^5 \partial_r \Psi\right)
 + {\omega^2 r_0^4\cosh^2\gamma\over r^2
(1 - r_0^4/ r^4
)} \Psi - j(j+4) \Psi =0
\ee
To solve this equation, first we need to isolate the singularity at the
horizon. The wavefunction at the horizon behaves as $\Psi \sim (1- r_0^4/
r^4 )^{i\kappa}$. It is therefore convenient to define
\be
\Psi = A\; \left(1- {r_0^4\over r^4} \right)^{i\kappa} f(r)\quad\quad
\kappa = {\omega r_0\cosh\gamma \over 4} = {\omega\over 4\pi T_H}
\ee
where $T_H = {1\over \pi r_0^2\cosh\gamma}$ is the Hawking temperature.
Then Eq.~(\ref{eq99}) becomes
\be\label{ennia}
r^2 \; \left(1- {r_0^4\over r^4} \right) \; f'' + r\; \left[
5 - (1-2i\kappa)\; {r_0^4\over r^4} \right] f' - j(j+4) f
= - {4\omega^2r_0^4\cosh^2\gamma\over r^2} \; {1+ {r_0^2\over r^2} + {r_0^4\over 
r^4} \over
1+ {r_0^2\over r^2} } \; f
\ee
The function $f$ has a regular limit as $r_0\to 0$ (as expected, since
we have already isolated the singularity in the wavefunction). Neglecting
higher-order corrections, we set $r_0 = 0$ in Eq.~(\ref{ennia}). The
result is
\be
r^2 f'' + 5r f'
+ {\omega^2r_0^4\cosh^2\gamma\over r^2}\; f- j(j+4) f =0
\ee
(notice that we kept the $r_0^4\cosh^2\gamma$ term, which remains finite and
proportional to the D-brane charge $R^4$ (Eq.~(\ref{dbch}))
whose solution is
\be
f(r) = {1\over r^2} \; H_{j+2}^{(1)} \left( {\omega r_0^2\cosh\gamma\over r} 
\right)
\ee
In the large $r$ limit, we have
\be
\Psi \approx -i\; A\; {2^{j+2}\; (j+1)! \over r_0^{2j+4}
\omega^{j+2} \cosh^{j+2} \gamma} \; r^j
\ee
Comparing with the asymptotic form~(\ref{tria}), we obtain
\be
A = i\; {\omega^{2j+4} r_0^{2j+4} \cosh^{j+2} \gamma \over 4^{j+2} (j+1)! 
(j+2)!}
\ee
The absorption coefficient is
\be\label{eq106}
{\cal P} = 8\pi\kappa r_0^4 \; |A|^2 \; |f(r_0)|^2
\approx 8\pi \kappa |A|^2 \; \Big| H_{j+2}^{(1)} (4\kappa) \Big|^2
\ee
Since $\kappa$ is large, we can expand the Bessel function,
\be
H_{j+2}^{(1)} (4\kappa) = {(-i)^{j+5/2}\over \sqrt{2\pi\kappa}}
\left( 1 + {i\over 8\kappa} \; (j+5/2)(j+3/2) + \dots \right)
\ee
Using the Gamma function identities~(\ref{gfi}), we may write
\be\label{happ}
H_{j+2}^{(1)} (4\kappa) = {-i\over 2^{2i\kappa} (2\kappa)^{j+2}} \;
e^{4i\kappa}\; {\Gamma (j/2+5/4-2i\kappa)\, \Gamma(j/2+7/4-2i\kappa)\over
\Gamma (1-4i\kappa)} + \dots
\ee
Using the approximation~(\ref{happ}), after some algebra we find that
for frequencies large compared to the temperature, the absorption 
coefficient~(\ref{eq106}) becomes
\be
{\cal P} \approx
{8\pi\kappa \over ((j+1)! (j+2)!)^2}
\; {|\Gamma(j/2+5/4+2i\kappa)\Gamma(j/2+7/4+2i\kappa)|^2\over
|\Gamma(1+4i\kappa)|^2}\; \left( {\omega r_0\over 2} \right)^{2j+4}
\ee
in line with the results we obtained in the extremal case ($\Delta =0$,
Eq.~(\ref{absco})), but at {\em half}~the Hawking temperature.

Comparing with the results in the near-extremal limit $\Delta \to 0$ 
(Eq.~(\ref{absne})),
we are led to the conclusion that $T_- \to\infty$, so that $T_+ = T_H/2$, in 
order
to satisfy
\be
{1\over T_+} + {1\over T_-} = {2\over T_H}
\ee
Thus, in this other end of the spectrum ($\Delta = 1$), it appears that one set
of fields in the conformal field theory (open string modes) live in a very hot
bath and decouple. Only one group survives.

In Fig.~1, we have plotted the dependence of the two temperature parameters
(for a given value of the Hawking temperature)
on the parameter $\Delta$, which is a measure of the departure from extremality.
The dotted lines indicate that we did not have an explicit result, but 
extrapolated
from the region where an explicit result was found.
In the extremal limit ($\Delta =0$), there is only one temperature parameter 
(the Hawking
temperature). At the other end of the range of $\Delta$ ($\Delta =1$), one of
the temperature parameters diverges, whereas the other one converges to half
the Hawking temperature. On the conformal field theory side, this behavior seems 
to
indicate that fields (open string modes) are not in equilibrium with each
other. For a given Hawking temperature, there are distinct vacua corresponding
to different supergravity configurations. This seems to imply that the fields form two
separate systems at two distinct temperatures. Of course, all this needs to be
taken with a grain of salt, since away from extremality
supersymmetry is broken and there is no guarantee that a duality exists between
supergravity and superconformal field theory on D3-branes. Yet, it is
intriguing that similar results are obtained in both the extremal and the
non-extremal cases. It would be interesting to extend these results to more
general supergravity backgrounds and probes, aiming at a better understanding
of the interesting issue of the AdS/CFT correspondence.

\newpage

\begin{center}
\setlength{\unitlength}{0.240900pt}
\begin{picture}(1500,900)(0,0)
\footnotesize
\put(1300,750){\makebox(0,0)[r]{$1/T_+$}}
\put(1300,250){\makebox(0,0)[r]{$1/T_-$}}
\thicklines \path(88,496)(108,496)
\thicklines \path(1433,496)(1413,496)
\put(66,496){\makebox(0,0)[r]{$1/T_H$}}
\thicklines \path(88,856)(108,856)
\thicklines \path(1433,856)(1413,856)
\put(66,856){\makebox(0,0)[r]{$2/T_H$}}
\thicklines \path(88,135)(88,155)
\thicklines \path(88,856)(88,836)
\put(88,90){\makebox(0,0){0}}
\thicklines \path(1433,135)(1433,155)
\thicklines \path(1433,856)(1433,836)
\put(1433,90){\makebox(0,0){1}}
\thicklines \path(88,135)(1433,135)(1433,856)(88,856)(88,135)
\put(760,73){\makebox(0,0){$\Delta$}}
\put(760,-100){\makebox(0,0){FIGURE 1: Dependence of the two D-brane
temperature parameters,
$T_\pm$, on the parameter $\Delta$}}
\put(760,-175){\makebox(0,0){{\phantom{FIGURE 1:1}} for a given Hawking temperature,
$T_H$. Explicit
results were derived in the small $\Delta$}}
\put(760,-250){\makebox(0,0){{\phantom{F}} region (near extremality) as well as
at $\Delta =1$ (non-rotating D-branes).}}
\Thicklines \path(88,496)(88,496)(102,522)(115,534)(129,543)(142,551)(156,558)(170,565)(183,571)(197,576)(210,582)(224,587)(237,592)(251,597)(265,602)(278,606)(292,611)(305,615)(319,619)(333,623)(346,627)
\Thicklines \dottedline[$\cdot$]{10}(360,631)(373,635)(387,639)(400,643)(414,647)(428,650)(441,654)(455,657)(468,661)(482,664)(496,668)(509,671)(523,675)(536,678)(550,681)(564,684)(577,688)(591,691)(604,694)(618,697)(631,700)(645,703)(659,706)(672,709)(686,712)(699,715)(713,718)(727,721)(740,724)(754,727)
\Thicklines \dottedline[$\cdot$]{10}(754,727)(767,730)(781,733)(794,736)(808,739)(822,741)(835,744)(849,747)(862,750)(876,752)(890,755)(903,758)(917,761)(930,763)(944,766)(957,769)(971,771)(985,774)(998,777)(1012,779)(1025,782)(1039,785)(1053,787)(1066,790)(1080,792)(1093,795)(1107,797)(1121,800)(1134,802)(1148,805)(1161,807)(1175,810)(1188,812)(1202,815)(1216,817)(1229,820)(1243,822)(1256,825)(1270,827)(1284,830)(1297,832)(1311,835)(1324,837)(1338,839)(1351,842)(1365,844)(1379,847)(1392,849)(1406,851)(1419,854)(1433,856)
\Thicklines \path(88,496)(88,496)(102,469)(115,457)(129,448)(142,440)(156,433)(170,426)(183,420)(197,415)(210,409)(224,404)(237,399)(251,394)(265,389)(278,385)(292,380)(305,376)(319,372)(333,368)(346,364)
\Thicklines \dottedline[$\cdot$]{10}(360,360)(373,356)(387,352)(400,348)(414,344)(428,341)(441,337)(455,334)(468,330)(482,327)(496,323)(509,320)(523,316)(536,313)(550,310)(564,307)(577,303)(591,300)(604,297)(618,294)(631,291)(645,288)(659,285)(672,282)(686,279)(699,276)(713,273)(727,270)(740,267)(754,264)
\Thicklines \dottedline[$\cdot$]{10}(754,264)(767,261)(781,258)(794,255)(808,252)(822,250)(835,247)(849,244)(862,241)(876,239)(890,236)(903,233)(917,230)(930,228)(944,225)(957,222)(971,220)(985,217)(998,214)(1012,212)(1025,209)(1039,206)(1053,204)(1066,201)(1080,199)(1093,196)(1107,194)(1121,191)(1134,189)(1148,186)(1161,184)(1175,181)(1188,179)(1202,176)(1216,174)(1229,171)(1243,169)(1256,166)(1270,164)(1284,161)(1297,159)(1311,156)(1324,154)(1338,152)(1351,149)(1365,147)(1379,144)(1392,142)(1406,140)(1419,137)(1433,135)
\Thicklines \dashline{30}(88,496)(1433,496)
\end{picture}
\end{center}\end{document}